\documentclass[remotesensing,article,accept,pdftex,moreauthors]{Definitions/mdpi}

\firstpage{1} 
\pubvolume{17}
\issuenum{3}
\articlenumber{352}
\pubyear{2025}
\copyrightyear{2025}
\externaleditor{Jose Moreno}
\datereceived{12 December 2024} 
\daterevised{10 January 2025}
\dateaccepted{17 January 2025} 
\datepublished{21 January 2025} 
\hreflink{https://doi.org/}

\usepackage{enumitem}
 
\Title{Development of Radar and Optical Tracking of Near-Earth Asteroids at the University of Tasmania}

\TitleCitation{Development of Radar and Optical Tracking of Near-Earth Asteroids at the University of Tasmania}

\Author{Oliver James {White} $^{1,}$*\orcidA{}, Guifré {Molera Calvés} $^{1}$\orcidB{}, Shinji {Horiuchi} $^{2}$\orcidC{}, Phil {Edwards} $^{3}$, Ed {Kruzins }$^{3,4}$,\linebreak Jon {Giorgini} $^{5}$, Nick {Stacy} $^{6}$, Andrew {Cole} $^{1}$, Chris {Phillips} $^{3}$, Jamie {Stevens} $^{3}$, Lance {Benner} $^{5}$ and Edwin {Peters} $^{4}$}

\AuthorNames{Oliver James {White}, Guifré {Molera Calvés}, Shinji {Horiuchi}, Phil {Edwards}, Ed {Kruzins }, Jon {Giorgini}, Nick {Stacy}, Andrew {Cole}, Chris {Phillips}, Jamie {Stevens}, Lance {Benner}, Edwin {Peters}}

\AuthorCitation{White, O.J.; Calvés, G.M.; Horiuchi, S.; Edwards, P.; Kruzins, E.; Giorgini, J.; Stacy, N.; Cole, A.; Phillips, C.; Stevens, J.; et al.}

\address{%
$^{1}$ \quad School of Natural Sciences, University of Tasmania, Hobart, TAS 7000, Australia; guifre.moleracalves@utas.edu.au (G.M.C.); andrew.cole@utas.edu.au (A.C.) \\
$^{2}$ \quad Commonwealth Scientific Industrial Research Organisation / NASA Canberra Deep Space Communication Complex, Tidbinbilla, ACT 2620, Australia; \linebreak shinji.horiuchi@csiro.au (S.H.) \\
$^{3}$ \quad Commonwealth Scientific Industrial Research Organisation, Sydney, NSW 2122, 
 Australia; \linebreak philip.edwards@csiro.au (P.E.); chris.phillips@csiro.au (C.P.); jamie.stevens@csiro.au (J.S.)\\
$^{4}$ \quad School of Electrical Engineering, University of New South Wales Canberra, Canberra 2610, Australia; e.kruzins@unsw.edu.au (E.K.); edwin.peters@unsw.edu.au (E.P.)\\
$^{5}$ \quad Jet Propulsion Laboratory, California Institute of Technology, Pasadena, CA 91011, USA; \linebreak jon.d.giorgini@jpl.nasa.gov (J.G.); lance.benner@jpl.nasa.gov (L.B.)\\
$^{6}$ \quad Independent researcher, Adelaide, Australia; nick.stacy@adam.com.au 
}

\corres{Correspondence: oliver.white@utas.edu.au}

\abstract{We detail the use of the University of Tasmania's (UTAS) optical and radio telescopes to conduct observations of near-Earth asteroids from 2021 to 2024. The Canberra Deep Space Communication Complex transmitted a radio signal at 7159.45 MHz, with the radar echo detected by the UTAS radio telescopes. The method of accounting for the Doppler shift between the stations and the near-Earth object is described so that others can implement a similar program. We present our results, with confirmed detections of 1994 PC1 and 2003 UC20 asteroids using the Hobart and Katherine 12-metre antennas, demonstrating the feasibility of using small radio telescopes for these observations. Additionally, the recently upgraded Ceduna 30\,m antenna was used to detect 2024 MK. Data collected from other observatories, such as Tidbinbilla, as well as the UTAS radar tracking of the moon are also presented in the context of demonstrating the means of applying these Doppler corrections and the accuracy of each method. Optical observations conducted in this period are also detailed as they complement radar observations and aid in refining the orbit parameters.}


\begin{document}


\section{Introduction}\label{Introduction}

Radar has been used to study Solar System objects for many decades in the northern hemisphere. As reviewed by \citet{Ostro_Planetary_Radar}, and more recently by \citet{Virkki}, radar observations of the Moon were first conducted post-World War II, and radar echoes had been detected from Mercury, Venus and Mars by 1963. Subsequent radar observations in the 1960s permitted the refinement of ephemerides to a sufficient accuracy to guide interplanetary spacecraft. Detections of Jupiter's Galilean moons, Saturn's rings and Titan were achieved in the 1970s and 1980s, as were the first echoes received from comets and asteroids. These have facilitated many developments in planetary science, including topographical features, orbital motion and surface composition. Additionally, as asteroids are remnants of the early Solar System, these studies can aid in developing an understanding of the Solar System's history.

Of particular interest are the observations conducted of near-Earth asteroids (NEAs), which have been conducted in the northern hemisphere since 1968 with the detection of Icarus (1566)~\cite{goldstein_icarus_1968}. These observations have scientific and planetary defence applications. Radar astrometry can be used to recover and refine known orbits, often by multiple orders of magnitude greater than by just considering optical observations~\cite{Ostro_Planetary_Radar}. The ramifications of this include improving spacecraft navigation when undertaking missions to these objects. For instance, 2016 HO3, a small, rapidly rotating NEA, is the target of the Tianwen-2 mission~\cite{2016HO3_2022}. Additionally, these observations can assist with planetary defence in the context of better predicting close approaches to Earth~\cite{belton_role_2004}.

The threat to life on Earth by an asteroid impact resulted in an increase in funding of these radar observations around the start of the 21st century. As a result, the number of these observations has increased tenfold from less than 100 to more than 1000 over the 20-year period from 2001 to 2021. Radar orbit determinations have confirmed that there are no known imminent threats of near-Earth objects (NEOs) impacting the Earth. A notable example is Apophis (99,942), a 340\,m asteroid discovered in 2005 that was thought to have a chance of impacting the Earth on 13 April 2029. However, subsequent observations, including radar, have determined that it will pass the Earth safely at a distance of 40,000\,km \cite{Virkki, giorgini_predicting_2008}.

\textls[-20]{In addition to refining their orbital parameters, the echoes from bistatic radar observations can also give insight into the surface and rotational properties of near-Earth objects and binary formations. This also includes surface topography and composition. For example, during the 2018 close approach of 2003 SD220, \citet{Horiuchi_2003_SD220} detected radar echoes using the Australia Telescope Compact Array (ATCA). These suggested an elongation of 1:8, which was consistent with light curve data suggesting a period of 285 $\pm$ 5 h \cite{warner_near-earth_2016}. Additionally, Delay-Doppler images were produced using Goldstone and the Green Bank Telescope (GBT), as described by \citet{Rivera_2003_SD220}. These revealed the elongated shape of 2003 SD220, which was consistent with the radar echoes detected at ATCA and other optical and near-infrared observations. Additionally, \citet{bondarenko2019radar} commensally used the 32\,m Zelenchukskaya and Svetloe radio telescopes to detect the Goldstone signal and confirmed the elongated shape of 2003 SD220. Delay-Doppler images of asteroids, such as that of NEA 2000 DP107 by~\citet{margot_binary_2002}, can also be used to detect whether an asteroid has any satellites.}

Additionally, radar ranging observations can detect the Yarkovsky effect on asteroids. This effect is an anisotropic non-gravitational acceleration of the asteroid due to re-emission of solar radiation after a time delay, which can result in changes to the asteroid's orbit. \citet{chesley_direct_2003} confirmed the Yarkovsky effect for (6489) Golevka, and this allowed the estimation of various physical parameters, such as the bulk modulus. We have ignored the impacts of the Yarkovsky effect in this paper.

However, until 2015, these planetary radar observations were conducted primarily in the northern hemisphere. Arecibo, the Goldstone 70\,m Deep Space Station 14 (DSS-14) and 34\,m DSS-13 antennas and the Green Bank Telescope (GBT) were used extensively for these observations. As described in~\citet{naidu_capabilities_2016}, DSS-14 and Arecibo were, respectively, capable of detecting 131 and 253 NEAs in monostatic configurations in 2015. Using GBT as a bistatic receiver with DSS-14 increased the potential detections to 195 NEAs in the same year.

Recently, the work for a future European facility for ground-based radar observations of near-Earth asteroids was also presented by \citet{Pupillo}. There are many benefits to implementing these programs at different locations for the study of near-Earth objects, particularly in the southern hemisphere, as pointed out by~\citet{giorgini_improved_2009} and~\citet{naidu_capabilities_2016}. These improvements could include the following:

\begin{itemize}
 \item Reduced uncertainty in near-Earth-object orbits.
 \item Doubled detections, including a small fraction which are not visible from the northern hemisphere or those which have stronger signal-to-noise ratios (SNRs) from the southern hemisphere. Furthermore, Australia offers additional longitudinal coverage compared to Arecibo and Goldstone or when there are scheduling conflicts.
 \item Further characterisation, such as the diameter, mass, rotation and surface properties.
 \item Better estimates of the pole direction through sequential observations in the northern and southern hemispheres.
\end{itemize}

In 2015, the first radar observations were conducted as part of the Southern Hemisphere Asteroid Research Program (SHARP) \cite{Benson_SHARP}. This involved using the antennas at the Canberra Deep Space Communication Complex (CDSCC), primarily the transmitter of the Tidbinbilla 70\,m antenna (DSS-43, Ti), and then receiving the radar echoes at the Parkes Radio Telescope and ATCA. In 2015, using this bistatic configuration, an estimated 70 NEAs were capable of being detected \cite{naidu_capabilities_2016}. Since then, many near-Earth objects have been observed from the southern hemisphere, with many objects observed between 2015 and 2020, with diameters from 20\,m to 5000\,m and at distances of 0.1 to 18 Lunar Distances (LD) \cite{Benson_SHARP, benson_detection_2017, abu-shaban_asteroids_2018, Molyneux_sharp_2021, Horiuchi_2003_SD220, Kruzins_deep_space_debris}. These exercises demonstrate the capabilities of the southern hemisphere to meaningfully contribute to the study of near-Earth objects using bistatic radar.

\begin{figure}[H]

\includegraphics[width=0.43\textwidth]{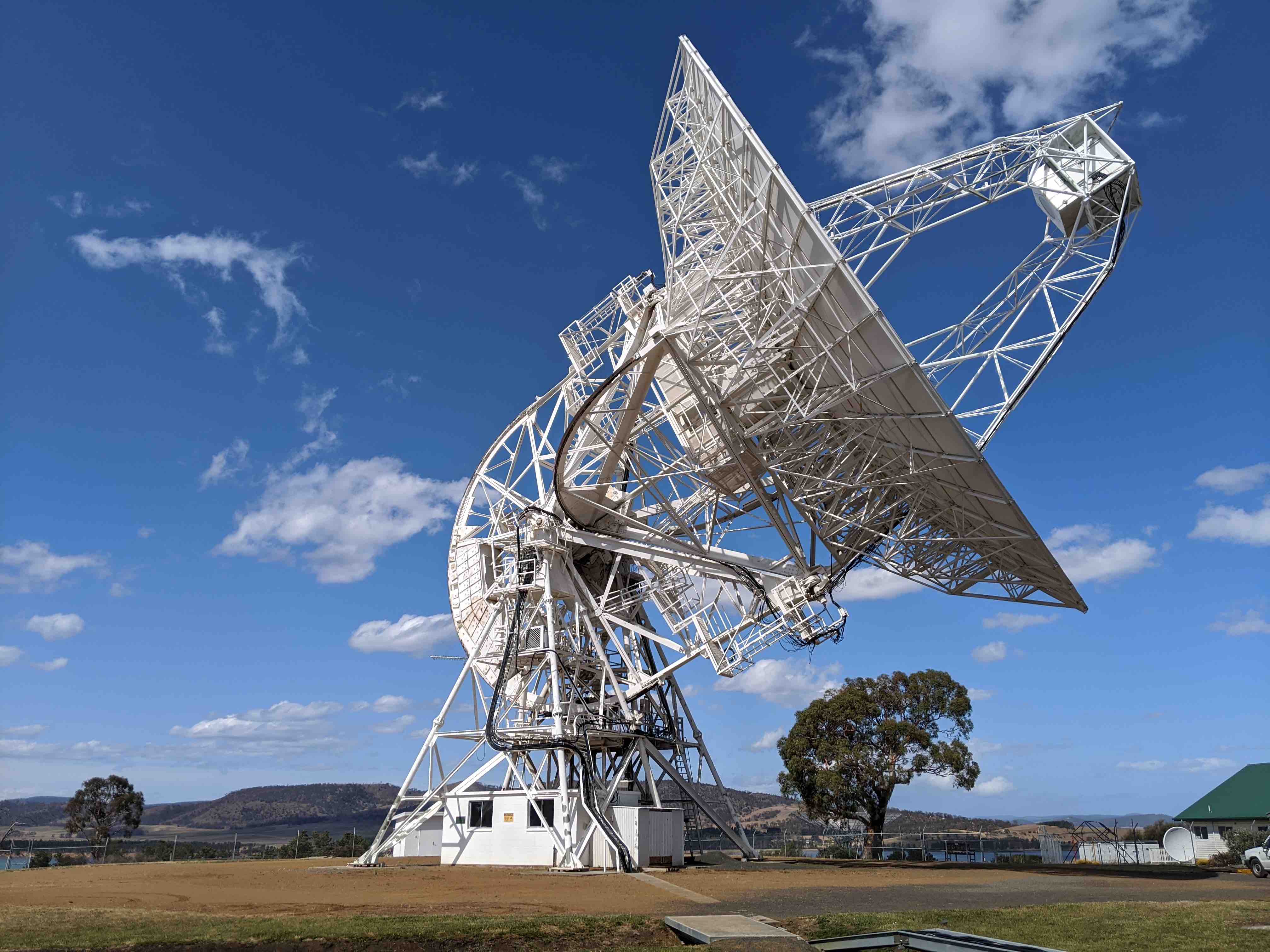}
\includegraphics[width=0.43\textwidth, trim={0 0 0 1.5cm}, clip]{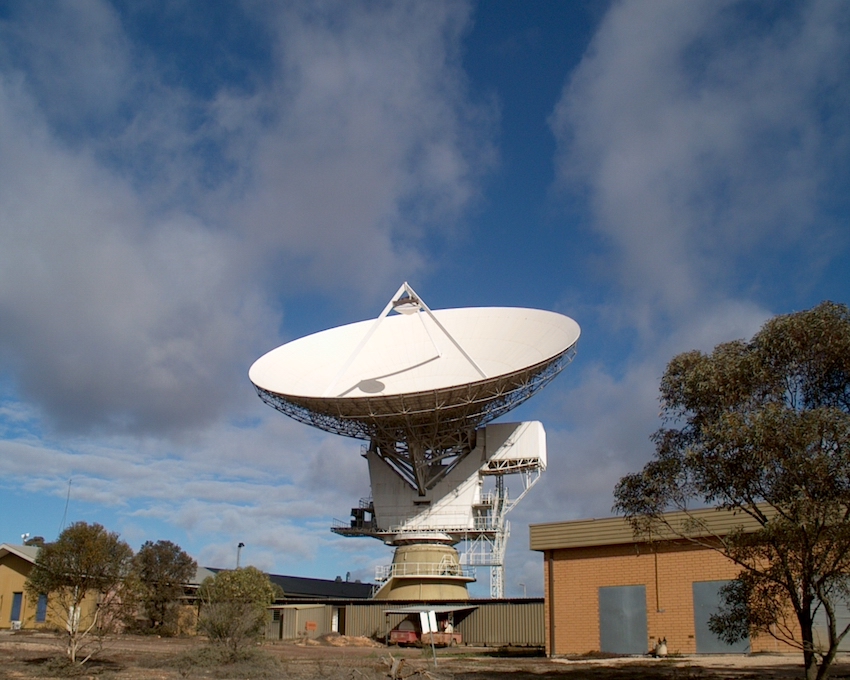} \\
\includegraphics[width=0.278\textwidth]{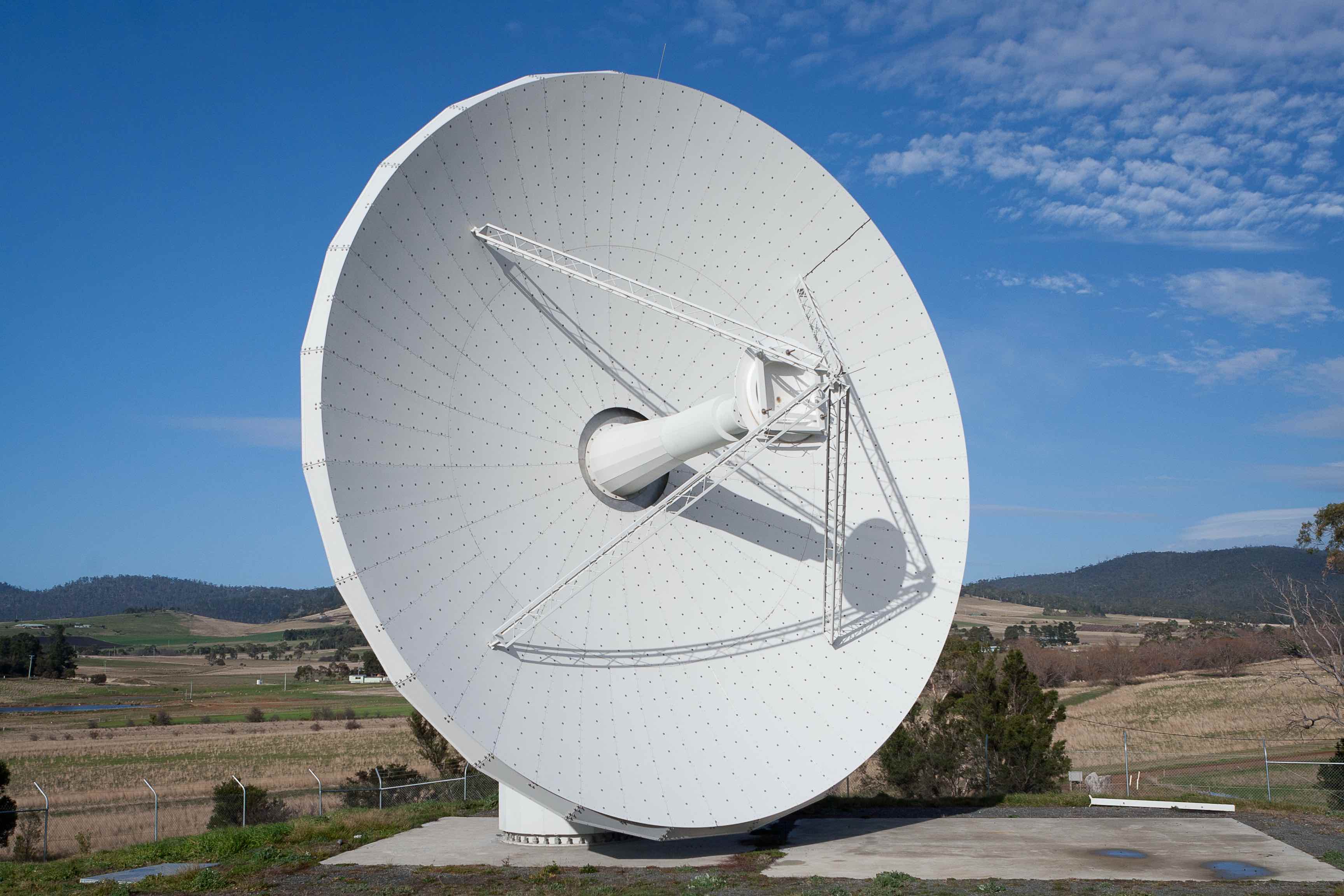}
\includegraphics[width=0.247\textwidth]{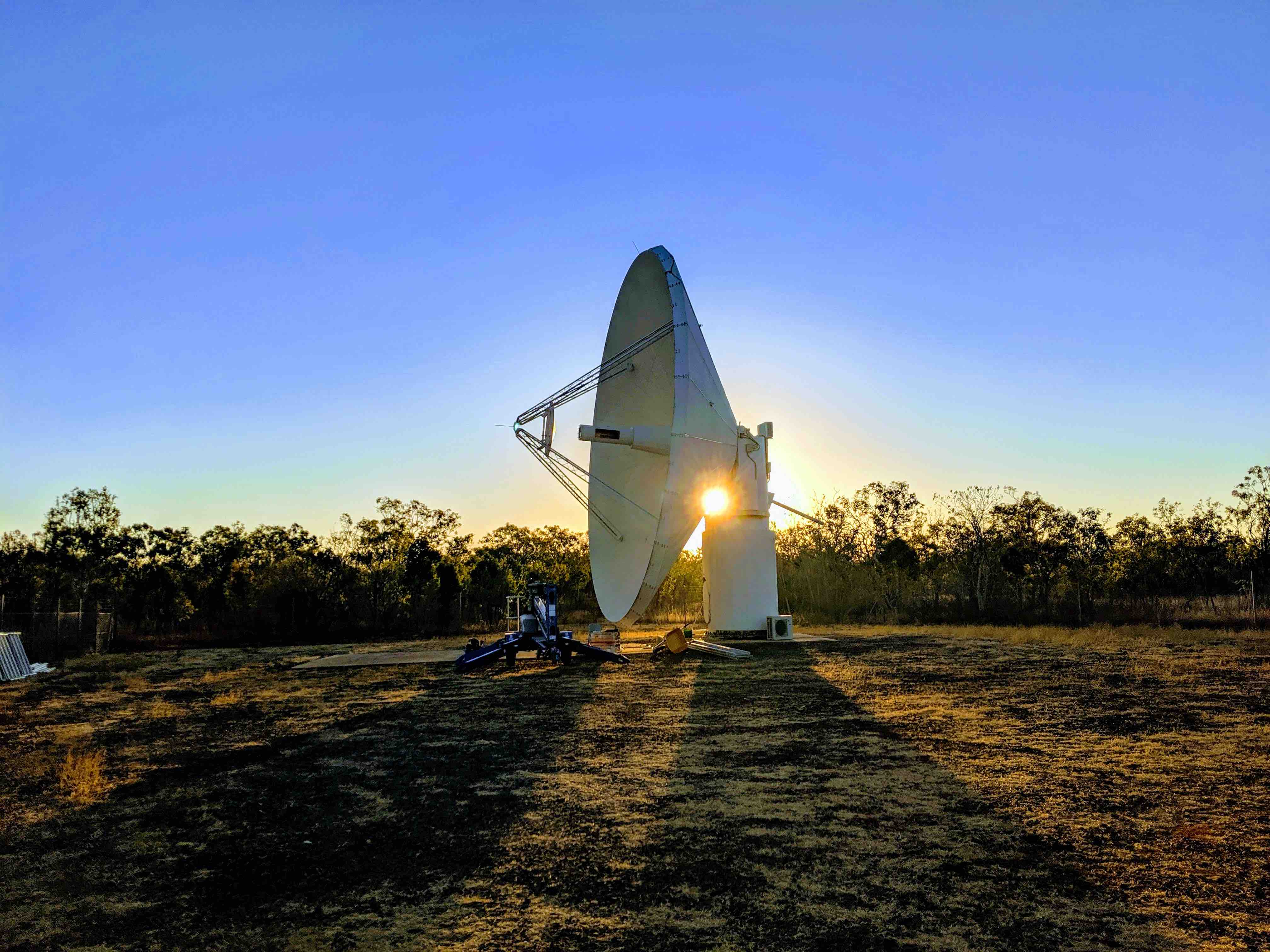}
\includegraphics[width=0.33\textwidth]{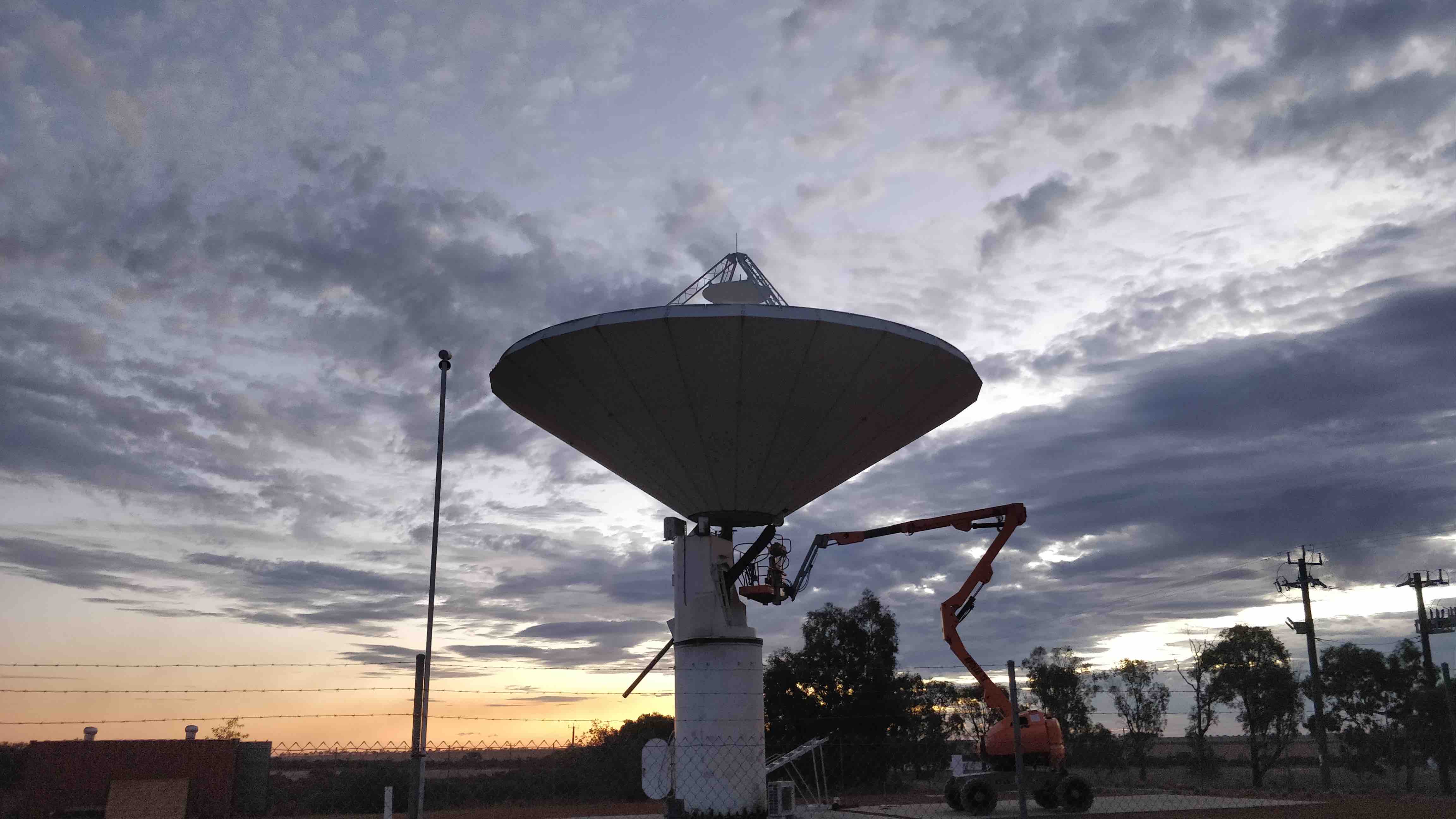}
\caption{UTAS radio telescopes. From top left in reading order are the Hobart 26\,m, Ceduna 30\,m, and the Hobart, Katherine and Yarragadee 12\,m antennas.}\label{Telescopes}
\end{figure}

The University of Tasmania (UTAS) joined SHARP in March 2021, contributing to the program through its continent-wide array of radio telescopes. This includes the 12\,m AuScope radio telescopes at Hobart (Hb), Katherine (Ke) and Yarragadee (Yg) \cite{lovell_auscope_2013}, the 26\,m at Hobart (Ho) and the 30\,m at Ceduna (Cd) \cite{mcculloch_cosmic_2005} (see Figure \ref{Telescopes}). These antennas add geographical diversity to the program, as well as the greater availability of receiving antennas. It also demonstrates the capacity of small antennas to contribute to these observations, as historically large radio telescopes such as Arecibo and Goldstone have primarily been used for the radar detection of near-Earth objects.

Furthermore, as discussed by \citet{Kruzins_deep_space_debris}, small aperture optical telescopes have participated in near-Earth-object observations in conjunction with SHARP. UTAS also contributes to this through its Greenhill Observatory at Bisdee Tier, which operates a 0.5\,m optical telescope. These contributions are also discussed here.

In this paper, we present the observations of near-Earth objects conducted by the University of Tasmania from 2021 to 2024 using both optical and bistatic radar techniques. The technical details of conducting the radar observations and the subsequent data processing are described in detail, with the intent of sharing our results as well as how others could implement a similar program. This demonstrates the capabilities of small antennas to take part in these bistatic radar observations. The ATCA results of the SHARP experiments during this period will be presented elsewhere, e.g., \citet{Kruzins_deep_space_debris, reddy_2023_2024} and Horiuchi et al. in preparation.

\section{Materials and Methods}\label{Methodology}

\subsection{Radar Observation Technique}\label{Radar Observation Technique}

\textls[-20]{The UTAS radio telescopes typically participate in these observations as receivers in a multistatic approach along with ATCA. This involves using the 20 kW transmitters at the Deep Space Stations (DSSs) at the CDSCC in Tidbinbilla, typically at the 70\,m DSS-43 or 34\,m DSS-36 antennas. The DSSs are capable of transmitting either left- or right-circularly polarised (L/RCP) electromagnetic continuous waves (CWs) in the C-band at approximately 7159.4 MHz. The demonstrations have also been conducted using the DSS-14 transmitter at the Goldstone Deep Space Communication Complex (GDSCC), California. DSS-14 is equipped with the Goldstone Solar System Radar (GSSR) transmitter, capable of transmitting a continuous wave or pulse modulated wave with a power of 450 kW at 8560.0 MHz. The common visibility between the UTAS antennas and Goldstone transmitter is low, and it only overlaps for almost an hour at very low elevation for targets near the celestial equator. Hence, we only conducted two experiments using Goldstone.}

Regardless of the transmitter used, the radio signal reflected from the near-Earth object is then received at the Australian antennas. The reflected signal contains both the same and opposite circular polarisations; however, the receivers at the UTAS 12\,m antennas are linearly polarised. Both Ceduna and Hobart-26 are equipped with circularly polarised C-band receivers.

Whatever frequency is transmitted will, however, be different at each receiver due to the Doppler shift, which can be approximately related (within tens of Hz) to the radial velocity between the stations and the near-Earth object:

\begin{equation}\label{eq:1}
f_D = \frac{(v_{tx - a} + v_{a - rx}) f_{tx}}{c_0}
\end{equation}

\noindent where $f_D$ is the Doppler frequency shift, $v_{tx - a}$ is the radial velocity between the transmitter and the NEO, $v_{a - rx}$ is the radial velocity between the NEO and receiver, $f_{tx}$ is the transmission frequency and $c_0$ is the speed of light. This is the Doppler shift associated with the bulk motion of the NEO’s centre of mass towards or away from the transmitter and receiver. In reality, contributions to the Doppler shift arise from general relativistic effects and gravitational effects from the bodies in the Solar System, as discussed in detail in~\citet{duev_spacecraft_2012,giorgini_site_2024}. Additionally, various sources of noise, such as frequency and timing noise, ionospheric plasma noise, unmodelled ground antenna motion, tropospheric scintillation, thermal and ground electronic noise and systematic effects, also play a role~\citep{armstrong_2006}. However, for the purposes of this paper, which aims to demonstrate the detection of NEOs, these factors are not further evaluated.

Of interest in the analysis of these radar observations is the Doppler shift due to the properties of the asteroid itself (e.g., rotation, surface properties). If all other effects can be neglected, then the Doppler broadening ($B$) of the radar echo is given by

\begin{equation}\label{eq:2}
B = \frac{4 \pi D(\phi) cos(\delta)}{\lambda P}
\end{equation}
where $D(\phi)$ is the diameter at rotation phase $\phi$, $\delta$ is the subradar latitude and $P$ is the rotation period. This is the Doppler broadening of the NEO echo due to relative rotation about the centre of mass, where the side of the asteroid moving towards the radar has a higher Doppler shift than the side moving away.

\begin{figure}[H]
\includegraphics[width=0.6\textwidth]{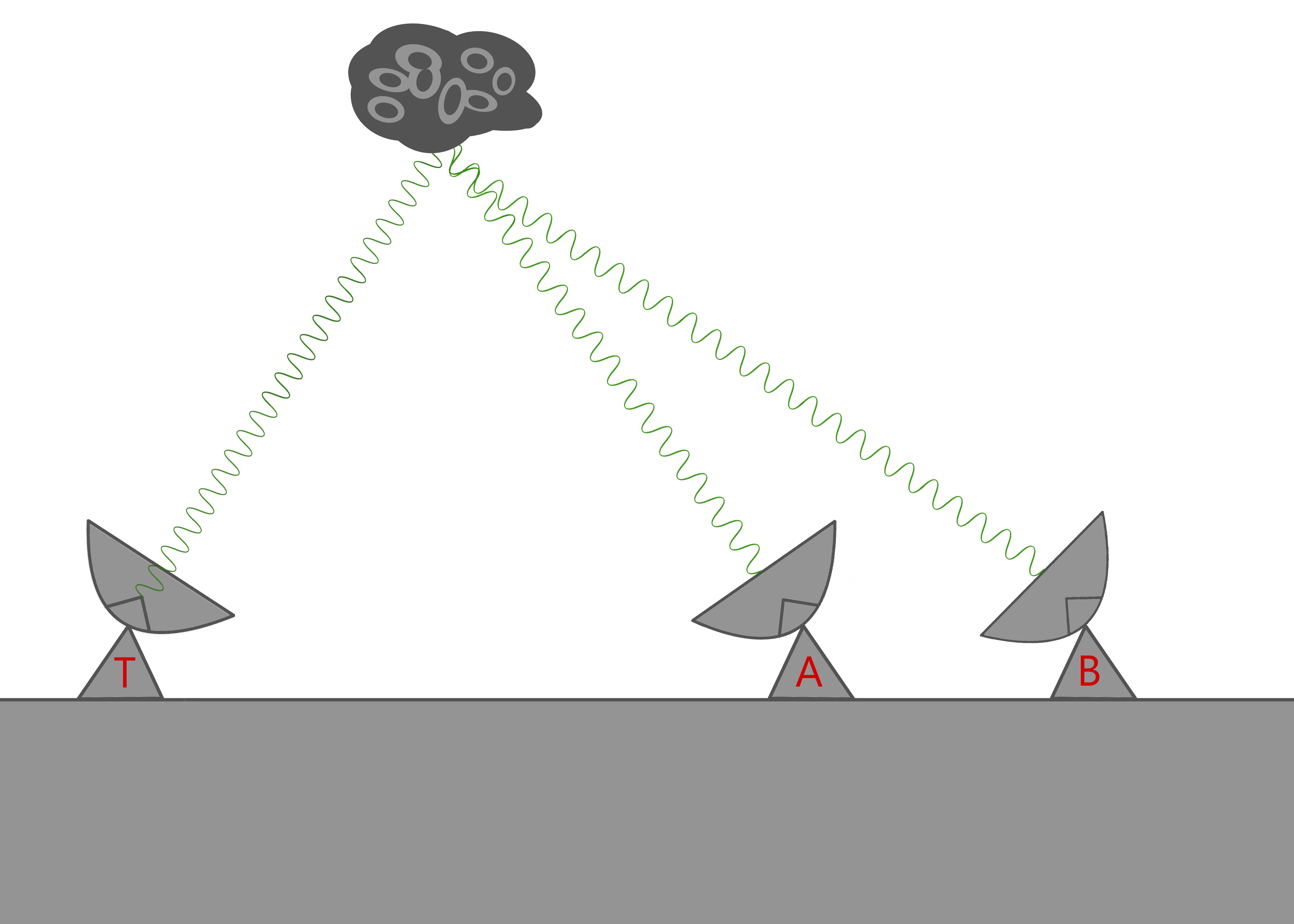}
\caption{Diagram of the bistatic radar setup. If antenna T transmits a static signal, then both receivers A and B will be required to perform the relevant Doppler compensation in the data processing due to the relative movement of the antennas and the asteroid. However, if antenna T transmits so that the signal is centred at receiver A, then B will be required to perform more complicated calculations.}\label{Bistatic_Radar_Diagram}
\end{figure}

\textls[-20]{Due to the relative motion of the transmitter at the CDSCC, the near-Earth object and the receiver, there will be additional contributions to the Doppler shift as given by Equation (\ref{eq:1}). These observations (see the schematics in Figure \ref{Bistatic_Radar_Diagram}) have been conducted in various ways, and thus require different methods of data processing to obtain the final signal:}

\begin{enumerate}[label=\alph*]
    \item A continuous fixed wave centred at 7159.45 MHz. The Doppler shift observed at the receivers includes the up- and down-leg contributions.
    \item A time-based shifted transmission frequency calculated to receive echoes at ATCA centred on 7159.45 MHz. This is the most common method of observation.
    \item A time-based shifted transmission frequency calculated to receive echoes at one of the UTAS antennas centred on 7159.45 MHz (typically Hobart).
    \item A monostatic dynamic tone or pulse centred for arrival at the same transmitting Goldstone antenna.
\end{enumerate}

\textls[-20]{Furthermore, there is the capacity for transmissions from multiple CDSCC antennas at different frequencies and/or polarisations to better facilitate detections from multiple receiving antennas. Typically, the CDSCC antennas transmit a right-circularly polarised radio wave, though there have been occasions where left-circular polarisation has been used.}
\vspace{-6pt}

\subsection{Experiment Setup}\label{Experiment setup}

The University of Tasmania operates five radio telescopes for astronomical, geodetic and spacecraft tracking purposes. These antennas are part of the Very Long Baseline Interferometry (VLBI) community, participating regularly in observations of the Australian Large Baseline Array (LBA) and of the International VLBI Service for Geodesy \& Astrometry (IVS). Therefore, all the sites are equipped with a standard VLBI back end, data acquisition systems and hydrogen-maser for the clock standards. Hobart-26 and Ceduna have narrow-band receivers covering standard astronomical VLBI frequency bands. The uplink transmitting frequencies from the DSSs are usually outside of the range. These include the S-band (2.02 GHz) and C-band (7.159 GHz) uplinks.

\begin{figure}[H]
\includegraphics[width=0.37\textwidth]{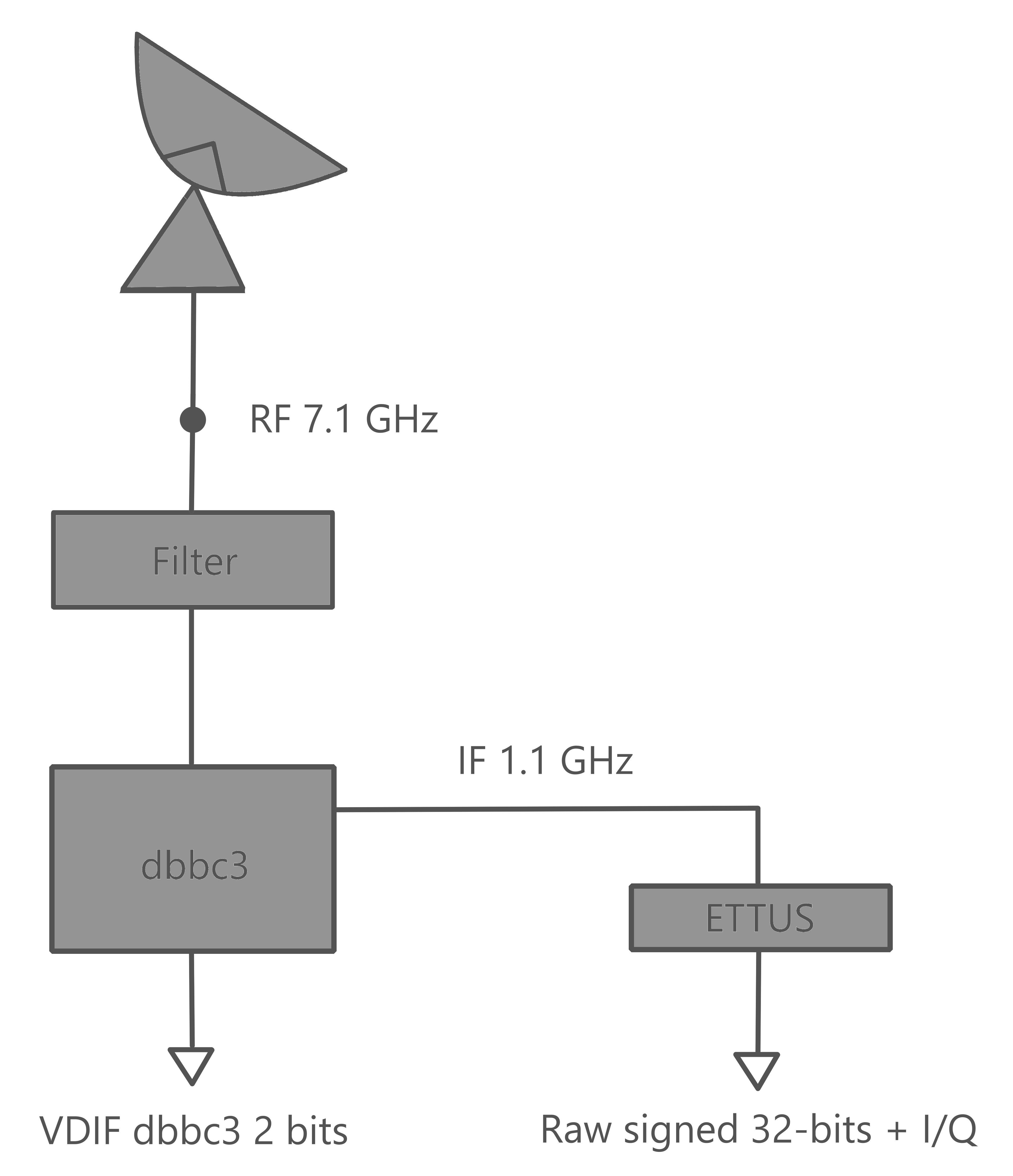}
\includegraphics[width=0.45\textwidth]{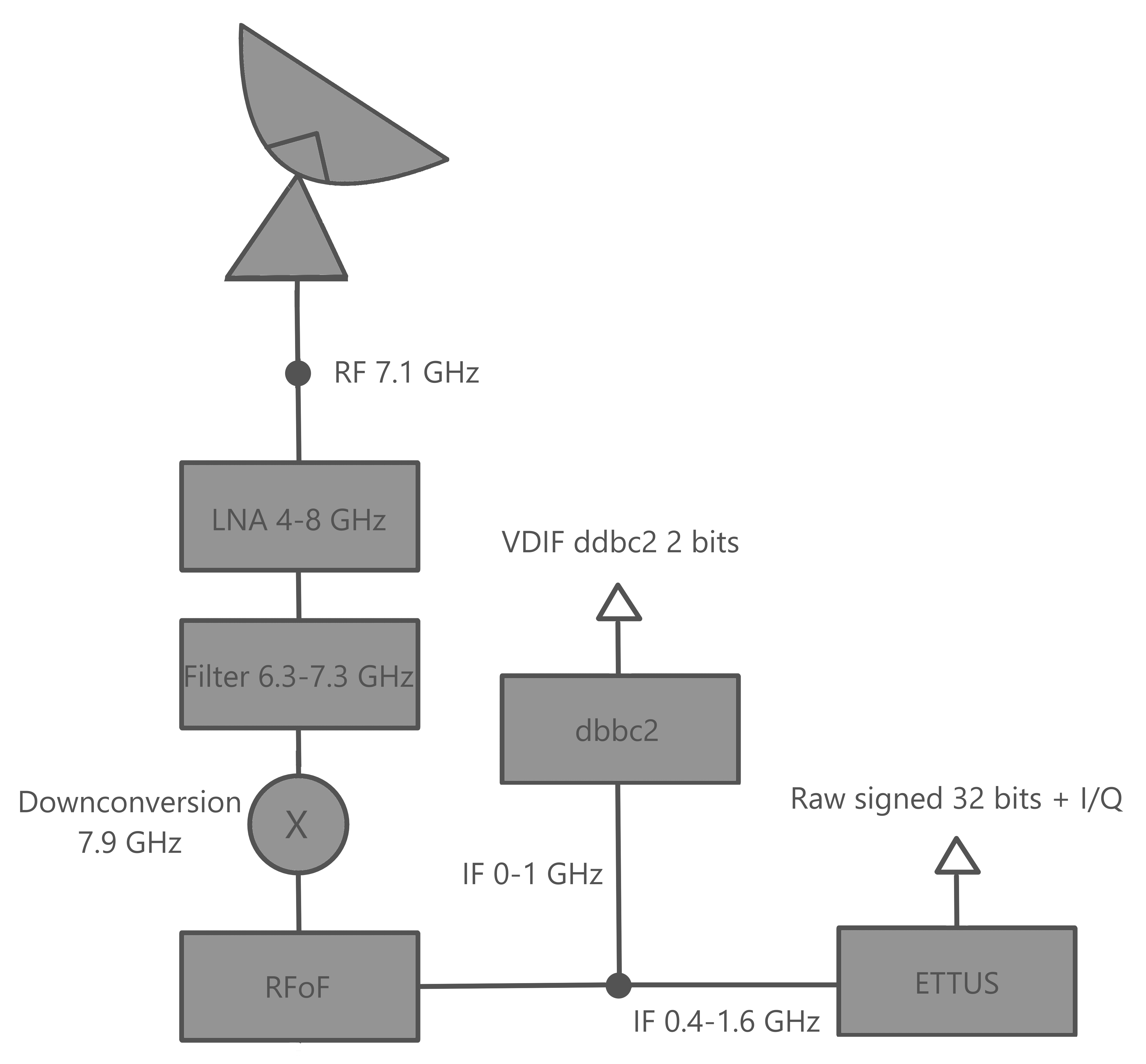}
\caption{Diagram of the data recording setup on the Hobart and Katherine 12\,m antennas (\textbf{left}) and Ceduna (\textbf{right}) as given in Section \ref{Experiment setup}.}
\label{Data_Recording}
\end{figure}

An upgrade to the Ceduna receiver in May 2024 made it possible for the antenna to join the SHARP observations. The upgraded receiver has a new wide-band Low-Noise Amplifier (LNA) covering the 4 to 8 GHz frequency range. In addition, new filters for the C-band receiver allows observations from approximately 6.3 GHz up to 7.3 GHz. Finally, new radio-frequency-over-fibre (RFoF) units were installed to minimise the signal chain losses from the receiver to the back end. The schematics of this design are shown in Figure \ref{Data_Recording}. A similar upgrade to the Hobart-26 antenna is being prepared for later in the year 2024.

The 12\,m antennas recently upgraded their receivers to a broadband system simultaneously covering any frequency between 3.0 and 14.0 GHz. Hence, Hobart-12 and Katherine-12 were used in these experiments. The receivers produce signals in linear polarisation (horizontal and vertical), while the transmitted signal is either right- or left-circular polarised. The radio frequency received by the radio telescopes is down-converted, within the 6--10 GHz sky frequency band, to an intermediate frequency (IF) in the range of \linebreak 0--4 GHz. The local oscillator (LO) for the down-conversion is set to 6.00 GHz. The analogue signal is then sampled with the digital baseband converter 3 (dbbc3) units~\citep{Guifre_SDtracker}. Initial tests used a single channel per polarisation of 32 MHz and 2-bit output. The limitation was due to an earlier version of the dbbc3 firmware. Later experiments used a newer version of the firmware, where we could tune the system to produce variable bandwidth, down to 2 MHz, and also 8 bits per sample. Using a larger number of bits per sample results in the improvement of the dynamic range \cite{benson_detection_2017}. Data are then recorded into a standard data acquisition system for further data processing.

In parallel, we developed a firmware design in software-defined radio (SDR) and gnuradio {\url{http://www.gnuradio.org/}} (accessed on 31 May 2024) tools to generate complex data at a higher number of bits per sample. The design runs in an Ettus X310~{\url{https://kb.ettus.com/Knowledge_Base}} (accessed on 29 May 2024) at the Hobart-12 site and in an Ettus N210 at Katherine. These SDR units were purchased to add some flexibility in the operations at the different sites. The current design outputs both real and complex data with 32 bits per sample using a bandwidth of 1 MHz. All data are then stored to disks for further data processing (Section~\ref{Data Processing}). Figure~\ref{Data_Recording} shows the schematics of the current design at our radio telescope sites. We have a plan to equip all our radio telescope infrastructure with SDR units to facilitate planetary radar operations, space situational awareness monitoring and spacecraft tracking operations.

{The primary goal of this paper is to establish bistatic radar receiving capability of the University of Tasmania radio telescopes, and we have proved that those work for strong targets and can increase radar detection coverage, especially when other resources to receive radar echoes are not available. Although multistatic radar is interesting for future applications, e.g., speckle observations to determine spin axes of asteroids~\cite{Busch_radar_speckles} or the incoherent addition of Doppler spectra recorded at different facilities, those are beyond the scope of this paper.}

\subsection{Data Processing}\label{Data Processing}

We modified the Spacecraft Doppler tracking (SDtracker) software version 2.0.14 \citep{Guifre_SDtracker}, developed by the Joint Institute for VLBI in Europe (JIVE) for Doppler measurements of deep space missions~\citep{Molera2014}, to conduct the data processing for planetary radar data. A similar approach in the data processing aspect was taken by~\citet{Pupillo}. SDtracker is utilised to compensate for any Doppler shift due to the movements of the near-Earth object and the stations, producing a frequency spectrum as the output containing only the residual Doppler effects due to the near-Earth object itself. In particular the multi-tone spacecraft tracker (SCtracker) component is utilised.

\textls[-20]{This takes the raw data recorded by one of the UTAS antennas or any VLBI radio telescope in the appropriate format, such as VDIF (VLBI Data Interchange Format) \cite{whitney_vdif_2009}. Files containing polynomial coefficients for the phase and frequency Doppler shift must also be supplied, which differ depending on the transmission methods discussed in Section \ref{Radar Observation Technique}. A coherent and non-coherent integration can be performed to produce a final spectrum. Should the transmission frequency be shifted to be centred on 7159.45 MHz at the UTAS antenna, then no Doppler corrections are required. In that case, polynomial coefficients that are all zero can be used.}

In the cases where the transmission frequency is centred at ATCA or has not been shifted at all, then a polynomial must be calculated which calculates the expected received frequency due to the Doppler shift from the radial movement of the stations and the near-Earth object. To do this, we can utilise the JPL Horizons online Solar System data service through the Python Astroquery module~\citep{Astroquery}. A script was created by the authors which uses this method to find the radial velocity between the stations and the near-Earth object and then uses Equation (\ref{eq:1}) to calculate the corresponding Doppler shift. This has different options depending on whether the transmission tone is static or centred on another station. It then creates a polynomial from the expected received frequency, with the coefficients outputted into files to be passed into SCtracker. The order of the polynomial fit depends on the length of the full recorded scan and motion of the asteroid. In most cases, this varies between 4 and 6.

\textls[-20]{Alternatively, for most observations, we are supplied predict files which contain the expected received frequency at each station based on a high-precision relativistic Doppler correction currently accurate to the second order. {This is created using the Jet Propulsion Laboratory’s ``On-Site Orbit Determination'' (``osod'') tool, developed in 1994. See~\citet{giorgini_site_2024} for a summary of this software and~\citet{JPL_1974} for a discussion of the physics.} Another script which uses these expected arrival frequencies to create a polynomial fit as above was developed.}

Once SCtracker has been run, there will be a number of outputs as given by \citet{Guifre_SDtracker}. We can use these to create plots of the summed frequency spectrum around the expected radar echo, as well as waterfall plots of the frequency as a function of time. Additionally, a mask file can be specified, which will remove the time segments for which the CDSCC transmitter was turned off (due to overhead aircraft, etc.), reducing unnecessary noise.

Once the frequency spectrum is obtained, we can normalise these based on the noise floor and subtract this so that positive amplitudes correspond to a signal detection rather than noise. The amplitude scale is then scaled in terms of the standard deviation of the noise. This gives us a sigma value for the peak SNR and is consistent with how these data are presented in the literature, such as by \citet{Horiuchi_2003_SD220}.

\subsection{Signal Strength}\label{Signal Strength}

The expected strength of the received radar echo is governed by the well-known radar range equation as given by \citet{Ostro_Planetary_Radar}, which for a bistatic radar system is
\begin{equation}\label{eq:3}
P_r = \frac{P_t G_t G_r \lambda^2 \sigma}{(4\pi)^3 R_{up}^2 R_{dw}^2}
\end{equation}
where $P_r$ is the received power, $P_t$ is the transmitted power, $G_t$ and $G_r$ are the transmitter and receiver antenna gains, respectively, $\sigma$ is the radar cross section (RCS) of the near-Earth object and $R_{up}$ and $R_{dw}$ are the distance between the stations and the near-Earth object.

Thus, when considering the received signal, the signal-to-noise ratio will be given by the ratio of $P_r$ to the standard deviation of the noise. The power of the noise is $P_{noise} = k T_{sys} \Delta f$, where $k$ is Boltzmann's constant, $T_{sys}$ is the temperature of the system (receiving antenna) and $\Delta f$ is the frequency resolution. Therefore, the standard deviation of the noise is given as $\Delta P_{noise} = {P_{noise}}/{(\Delta f \Delta t)^{1/2}}$, where $\Delta t$ is the total integration time.

\textls[-20]{Therefore, the SNR of the received signal from the bistatic radar method will be given by}
\begin{equation}\label{eq:4}
SNR = \frac{P_t G_t G_r \lambda^2 \sigma (\Delta t)^{1/2}}{(4\pi)^3 R^4 k T_{sys} (\Delta f)^{1/2}}
\end{equation}

This equation can be used either to estimate the expected SNR for an observation if the RCS is known or estimated (e.g., a spherical approximation) or, when a detection occurs, to calculate the RCS of the near-Earth object, as all other terms should be known.

\subsection{Observations}\label{Observations}

From March 2021 to June 2024, there were over 40 separate SHARP observations of different near-Earth objects in which the UTAS radio telescopes participated. These involved the Hobart, Katherine and Yarragadee 12 m as well as the Ceduna 30\,m and Hobart 26\,m radio telescopes. See Table \ref{Observations Table} for a summary of these. ATCA also participated as a receiver in all bar one of these observations. There were detections at the UTAS antennas for the 1994 PC1, 2003 UC20, Moon and 2024 MK observations. The observations for which the CDSCC antennas acted as transmitters used a continuous wave radio signal, so radar ranging techniques could not be performed.

Calculating the expected SNR using Equation (\ref{eq:4}), the majority of the non-detections with the UTAS 12\,m antennas were not unexpected, with expected signal strengths weaker or not much larger than the noise floor. However, an SNR of 7 and 12 dB was expected for the 26/12/2022 and 27/12/2022 2010 XC15 observations, respectively. Equipment and recording errors at the UTAS antennas resulted in the non-detection for this experiment.

Furthermore, the optical telescopes at Bisdee Tier conducted one successful observation of a near-Earth object in conjunction with the bistatic radar, which is described further in Section \ref{Optical Detections}.

\begin{table}[H]
\scriptsize
    \caption{Bistatic radar observations of near-Earth objects with the the University of Tasmania radio telescopes participating as receivers between between 2021 and 2024. Range and size values obtained from the NASA JPL Horizons service. At and Ut indicate ATCA and UTAS detections, respectively.}\label{Observations Table}
    \begin{tabularx}{\columnwidth}{p{0.18\textwidth}CCCCCC}
    \toprule
   \textbf{  Near-Earth Object} & \textbf{Epoch} & \textbf{Tx DSS Stations} & \textbf{Rx Stations} & \textbf{Range (LD)} & \textbf{Diameter (m)} & \textbf{Detection} \\
    \midrule
     2001 FO32 & 21/03/2021 & 34, 36, 43 & Hb, Ke & 5.2 & 1000 & At\\
     2021 AF8 & 05/05/2021, 06/05/2021 & 43 & Hb, Ke & 8.8 & 310 & At\\
     2021 GK1 & 11/05/2021 & 43 & Hb, Ke & 1.5 & 14 & No\\
     Nereus & 15/12/2021 & 43 & Hb, Ke & 11.4 & 610 & At\\
     2003 SD220 & 20/12/2021, 21/12/2021 & 43 & Hb, Ke, Cd & 14.4 & 1000 & At\\
     1994 PC1 / Moon & 18/01/2022 & 43 & Cd & 5.1 & 1600 & At+Ut\\
     1994 PC1 & 19/01/2022 & 36, 43 & Hb, Ke, Cd & 5.2 & 1600 & At+Ut \\
     2001 CB21 & 03/03/2022 & 36 & Ke & 12.8 & 680 & At\\
     2012 UX68 & 17/05/2022, 18/05/2022 & 35 & Hb, Ke & 3.3 & 50 & No\\
     1989 JA & 28/05/2022 & 36 & Hb, Ke & 10.5 & 990 & At\\
     2022 LV & 25/06/2022, 26/06/2022 & 14 & Hb, Ke & 2.0 & 22 & At\\
     2022 RM4 & 01/11/2022, 02/11/2022 & 43 & Hb, Ke & 6.0 & 410 & At\\
     2005 LW3 & 22/11/2022 & 43 & Hb, Ke & 3.5 & 400 & At\\
     2015 RN35 & 14/12/2022 & 43 & Hb, Ke & 1.9 & 75 & At\\
     2014 HK129 & 19/12/2022 & 14 & Ho & 6.7 & 210 & At\\
     2010 XC15 & 26/12/2022, 27/12/2022 & 34 & Hb, Ke & 2.0 & 180 & At\\
     2011 AG5 & 02/02/2023, 03/02/2023 & 43 & Hb, Ke & 4.7 & 150 & At\\
     2005 YY128 & 15/02/2023 & 43 & Hb, Ke & 12.0 & 740 & No\\
     2012 KY3 & 13/04/2023 & 35 & Hb & 12.8 & 710 & At\\
     2006 HV5 & 28/04/2023 & 14 & Ho & 9.5 & 450 & At\\
     1994 XD & 09/06/2023, 10/06/2023 & 34, 14 & Hb, Ho & 9.6 & 490 & No\\
     2018 UY & 11/07/2023, 12/07/2023 & 43 & Hb, Ke & 7.4 & 250 & At\\
     2020 UQ3 & 15/07/2023, 16/07/2023 & 34 & Hb, Ke & 4.7 & 58 & At\\
     2016 LY48 & 16/09/2023, 17/09/2023 & 36, 43 & Hb & 5.0 & 96 & At\\
     1998 HH49 & 16/10/2023, 17/10/2023 & 36 & Hb, Ke & 3.0 & 190 & At\\
     2003 UC20 & 04/11/2023 & 43 & Hb & 13.8 & 680 & At+Ut\\
     Moon & 13/06/2024 & 43 & Cd, Hb, Ke & 1.0 & 3,474,800 & Ut \\
     2011 UL21 & 26/06/2024, 28/06/2024 & 43 & Cd & 17.3 & 2200 & At\\
     2024 MK & 26/06/2024, 28/06/2024, 29/06/2024 & 43, 35 & Cd, Hb, Ke, Yg & 0.8 & 140 & At+Ut \\
    \bottomrule
    \end{tabularx}
\end{table}

\section{Results}\label{Results}

The results section is organised thematically rather than chronologically. We utilised various experiments to illustrate the different methodologies employed to successfully detect radar echoes from asteroids.

\subsection{Moon Detection}\label{Moon Detection}

\begin{figure}[H]
 \includegraphics[width=0.85\textwidth]{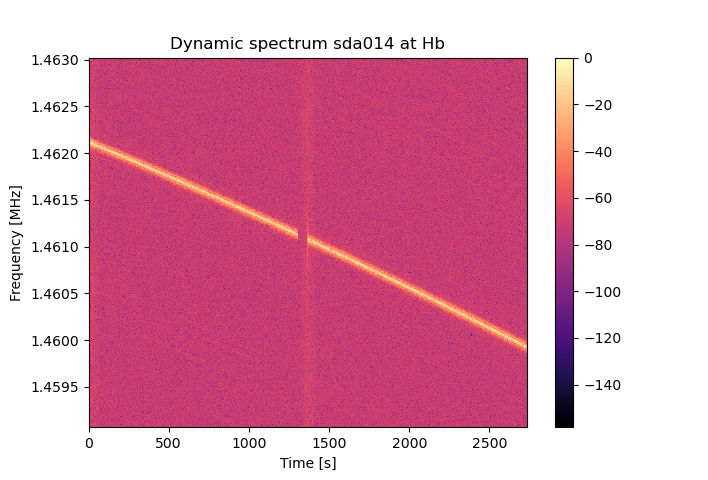}
 \caption{Waterfall spectra plot of the bistatic radar observation of the Moon from Hb with DSS-43 transmitting on 13 June 2024. No Doppler compensation has been applied. Total observation length is 45 min.}
 \label{Moon_NoDop}
\end{figure}

In order to test the upgraded Ceduna receiver, we conducted bistatic radar experiments of the Moon. On 13 June 2024, we used the DSS-43 antenna to transmit a signal towards the Moon, and the reflected signal was observed by the Ceduna, Katherine and Hobart radio telescopes. The transmission frequency was constant at 7159.45 MHz with a power of 20 kW, and without Doppler compensation there was a drift of approximately 2000 Hz throughout the session. Figure \ref{Moon_NoDop} shows the waterfall plot of the spectra around the expected receiving frequency at Hb using the y linear polarisation. The y-axis is the frequency within the IF band. The transmission mode was switched from LCP to RCP partway through the observation for calibration purposes, as seen in the transmission break in the waterfall plot.

\begin{figure}[H]
\includegraphics[width=0.85\textwidth]{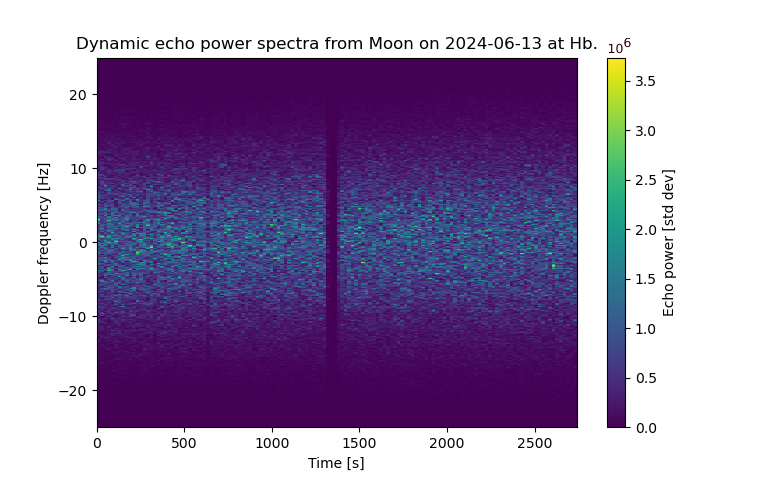}
\caption{\textls[-20]{Waterfall spectra plot of the bistatic radar observation of the Moon from Hb with DSS-43 transmitting. The {Doppler compensation using the instantaneous radial velocity approximation} was applied.}}\label{Moon_At_Horiz}
\end{figure}

Utilising the publicly available instantaneous radial velocity to calculate the polynomial fit and applying this to the received signal, the Doppler drift is reduced to negligible levels over the duration of the observation, as given by Figure~\ref{Moon_At_Horiz}. This demonstrates the usefulness of the instantaneous radial velocity approximation of Doppler for a static transmitted signal in this instance. The signal is centred on 0 Hz (expected arrival frequency).

\begin{figure}[H]
\includegraphics[width=0.85\textwidth]{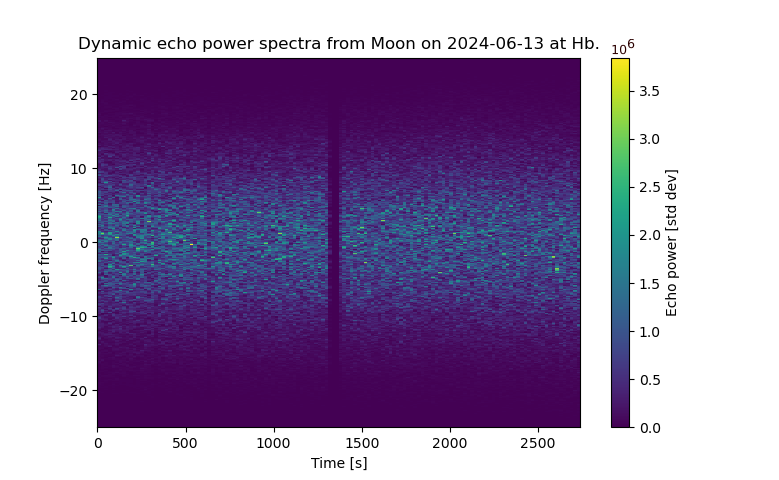}
\caption{\textls[-20]{Waterfall spectra plot of the bistatic radar observation of the Moon from Hb with DSS-43 transmitting. The spectra were used to perform the Doppler compensation, facilitated by the high SNR.}}\label{Moon_Spec}
\end{figure}

Due to its relative proximity and large radar cross section, the reflected signal is detectable on each individual integration. This high SNR (7.7$\times$10$^{\,6}$--sigma for the summed spectra) enables the use of each spectrum to directly compensate for the Doppler shift of the received frequency using the aforementioned SCtracker software (v2.0.14). In principal, the output of this has less drift than the instantaneous radial velocity approximation, though it is only applicable to targets of sufficient strength. See Figure \ref{Moon_Spec}, though for this observation there was no discernible difference between these two methods.

\subsection{2014 HK129}\label{2014 HK129}

On 27 April 2014, 2014 HK129 was discovered by the Catalina Sky Survey \citep{2014HK129_discovery}. Light curve observations conducted by the Ondrejov Asteroid Photometry Project indicate a rotation period of 15.119 h \citep{Ondrejov_period}. It is an Apollo-class NEO with an absolute magnitude of 21.10 with an estimated diameter of 160--360 m, and it underwent a close approach of Earth on 20 December 2022, with a nominal close approach distance of 0.017\,AU \citep{small_body_database_lookup}.

On 19 December 2022 and 22 December 2022, Goldstone was transmitting a monostatic radio wave at 8560 MHz, targeting the asteroid 2014 HK129. This involved transmitting pulses for a duration of twice the light travel time to 2014 HK129 and then stopping the transmission to receive the pulse. A receiver at Tidbinbilla (DSS-43) eavesdropped on this observation. We also attempted using Hobart-26 to eavesdrop on the GSSR signal, but the elevation of the target was too low for the antenna. Hobart-26's elevation limit is 10 degrees. On first inspection of the frequency spectrum at Tidbinbilla, without applying any additional Doppler correction to the data, the radar echo was not detected.

\begin{figure}[H]
\includegraphics[width=0.8\textwidth]{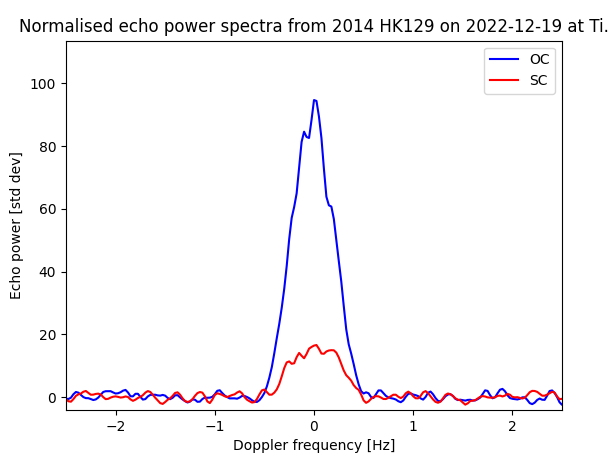}
\caption{Sum of spectra from the bistatic radar observation of 2014 HK129 from Tidbinbilla with Goldstone transmitting on 19 December 2022. The predict file method has been used for performing the Doppler compensation, and a frequency resolution of 0.05 Hz has been used. A strong radar echo was detected with an SNR of 95-sigma in the opposite circular polarisation and 17-sigma in the same circular polarisation. Total scan length is 92 min.}\label{2014HK129_Tid_19}
\end{figure}

\begin{figure}[H]
\includegraphics[width=0.8\textwidth]{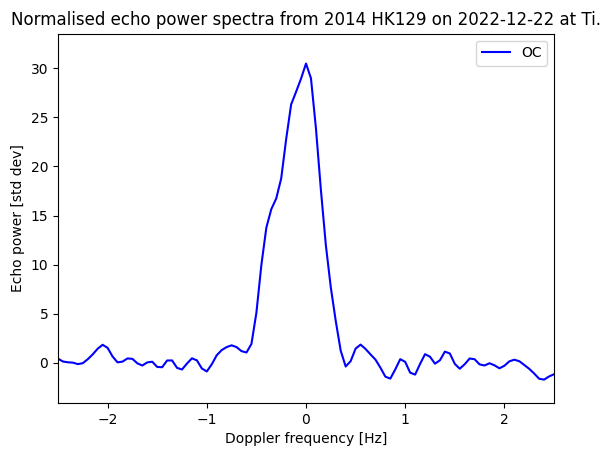}
\caption{Sum of spectra from the bistatic radar observation of 2014 HK129 from Tidbinbilla with Goldstone transmitting on 22 December 2022. The predict file method has was used to perform the Doppler compensation, and a frequency resolution of 0.1 Hz was used. A strong radar echo was detected with an SNR of 31-sigma in the opposite circular polarisation. Total scan length is 31 min.}\label{2014HK129_Tid_22}
\end{figure}

In this case, we used the predict files provided by NASA/JPL to create a polynomial fit in phase of order 4. This polynomial fit was used to compensate for the Doppler difference in the received signal. We experimented with multiple polynomial orders and with the least square method for the best fit. A polynomial of order 4 was sufficient to obtain a standard deviation of the expected Doppler noise better than 0.5 Hz. After Doppler compensation, we stacked all the resulting spectra in a single echo. The resulting spectrum has a singular peak from the detected echo, see Figures \ref{2014HK129_Tid_19} and \ref{2014HK129_Tid_22}. This was consistent with {Goldstone observations} {\url{https://echo.jpl.nasa.gov/asteroids/2010XC15/2010XC15.2022.goldstone.planning.html}} (accessed on 14 March 2024) conducted around the same time.

Based on its rotation period and diameter, the Doppler broadening of the radar echo as given in Equation (\ref{eq:2}) is 1--2 Hz (depending on the precise diameter of the body), which is consistent with the value observed at Tidbinbilla.

\newpage

\begin{figure}[H]
\includegraphics[width=0.85\textwidth]{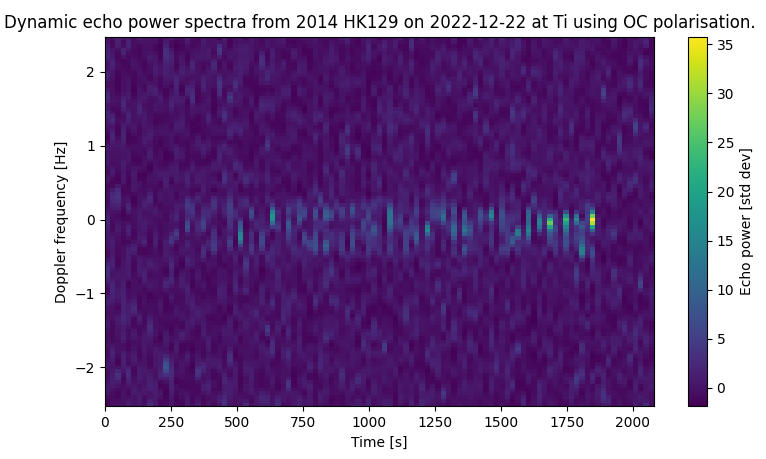}
\caption{Waterfall spectra plot from the bistatic radar observation of 2014 HK129 from Tidbinbilla with Goldstone transmitting on 22 December 2022 following Doppler compensation using the predict file method.}\label{2014HK129_Tid_22_Waterfall}
\end{figure}

Looking at the waterfall plot for these observations clearly shows the individual pulses transmitted from Goldstone. See Figure \ref{2014HK129_Tid_22_Waterfall} for the 22 December 2022 data. We were not able to process the data as given in Figures \ref{2014HK129_Tid_19} and \ref{2014HK129_Tid_22} to not include the gaps in transmission, which resulted in an increase in the noise floor. Automating this in the data processing pipeline is currently being worked on.

\subsection{1994 PC1}\label{1994 PC1}

1994 PC1 was discovered on 9 August 1994 by the Siding Spring Observatory \citep{1994PC1_discovery}. It has a rotation period of 2.6 hours as determined by the Ondrejov Asteroid Photometry Project \citep{1994PC1_period}, and NASA's Wide-field Infrared Survey Explorer (NEOWISE) mission found its diameter to be $1.052 \pm 0.303$ km \citep{1994PC1_diameter}. It is an Apollo-class NEO which had a close approach to Earth of 0.013\,AU on 18 January 2022 \citep{small_body_database_lookup}.

\begin{figure}[H]
\includegraphics[width=0.6\textwidth]{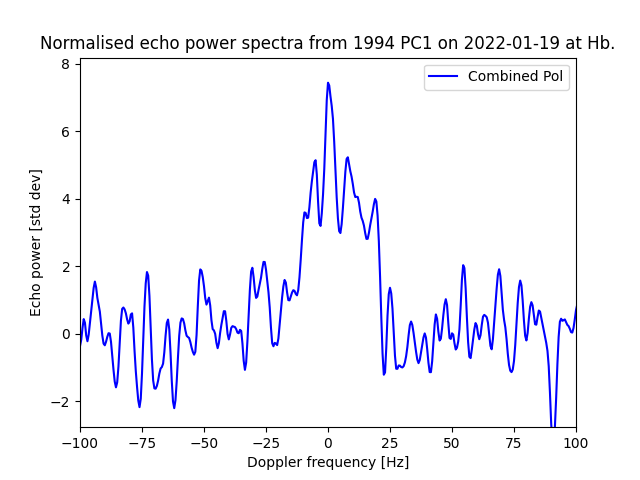}
\caption{Sum of spectra from the bistatic radar observation of 1994 PC1 from Hb on 19 January 2022 with DSS-43 transmitting. Net length of scan is 55 min, starting at 07:53 UTC. The frequency received at the zero point is 7159.45 MHz, with a frequency resolution of 1 Hz smoothed using a Savitzky--Golay filter \citep{savitzky_savgol_1964} with window size 10 and a polynomial fit of order 3. Radar echo detected with an SNR of 7-sigma when combining both linear polarisations.}\label{1994PC1_Hb}
\end{figure}

\begin{figure}[H]
\includegraphics[width=0.6\textwidth]{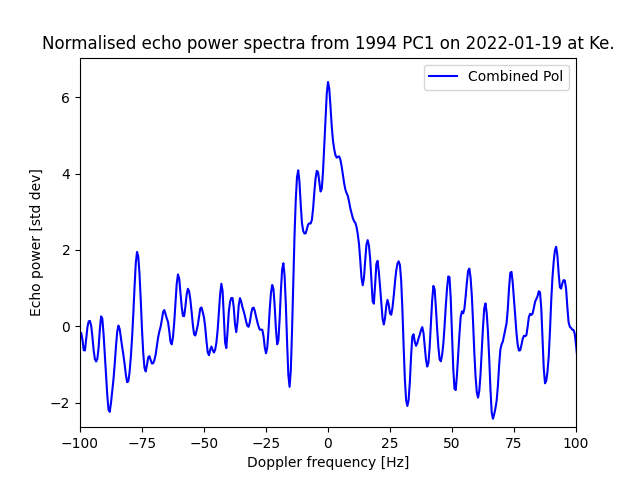}
\caption{Sum of spectra from the bistatic radar observation of 1994 PC1 from Ke on 19 January 2022 with DSS-43 transmitting. Net length of scan is 55 min, starting at 07:53 UTC. The frequency received at the zero point is 7159.45 MHz, with a frequency resolution of 1 Hz smoothed using a Savitzky--Golay filter \citep{savitzky_savgol_1964} with window size 10 and a polynomial fit of order 3. Radar echo detected with an SNR of 6-sigma when combining both linear polarisations.}\label{1994PC1_Ke}
\end{figure}

On 19 January 2022, the 12\,m antennas at Hobart and Katherine detected the radar echo from 1994 PC1 with DSS-43 transmitting. The SNRs at Hobart and Katherine were 7-sigma and 6-sigma, respectively, which demonstrates the ability of small radio telescopes to contribute to these observations. Furthermore, this observation had the transmission frequency centred for ATCA, and both rigorous Doppler and instantaneous radial velocity approximate compensation methods were independently utilised to produce the final spectra. Figures \ref{1994PC1_Hb} and \ref{1994PC1_Ke} show the integrated spectra of 1994 PC1 acquired with Hobart-12 and Katherine, respectively, using the instantaneous radial velocity compensation.

Based on its rotation period and diameter, the Doppler broadening of the radar echo as given in Equation (\ref{eq:2}) is 34 Hz (ignoring effects due to the non-sphericity of the body). This is comparable to the broadening observed at Hobart and Katherine.

\vspace{-10pt}

\begin{figure}[H]
\includegraphics[width=0.6\textwidth]{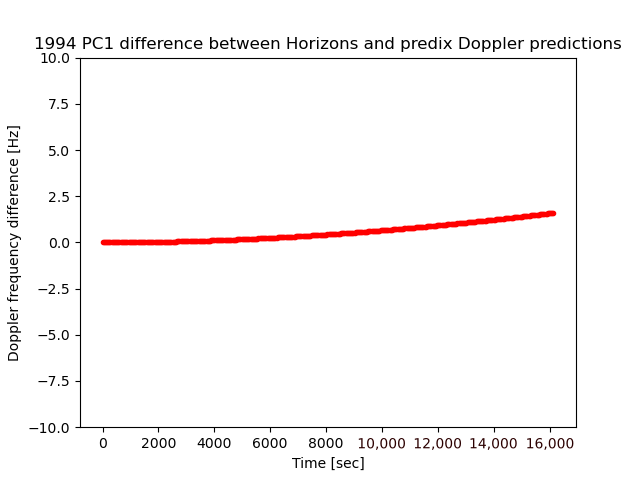}
\caption{Difference in Doppler predictions for asteroid 1994 PC1 at Hb on 19 January 2022. The estimates compare Doppler measurements with the On-Site Orbit Determination by JPL and the instantaneous radial velocity approximation.}\label{Doppler Difference}
\end{figure}

We employed the instantaneous radial velocity compensation to process the data from both stations given that the scan durations were under an hour. However, for longer observation tracks, it is essential to utilise the On-Site Orbit Determination Doppler files created by JPL. Figure~\ref{Doppler Difference} shows the Doppler difference obtained using both methods for asteroid 1994 PC1 on 19 January 2022 at Hobart. They are set arbitrarily to 0 offset at the beginning of the scan.
\vspace{-8pt}

\subsection{2003 UC20}\label{2003 UC20}

2003 UC20 is an Aten-class NEO discovered by the Lincoln Near-Earth Asteroid Research project on 21 October 2003 \citep{small_body_database_lookup}. The Ondrejov Asteroid Photometry Project determined its period to be 29.6 h, albeit with a large uncertainty \citep{Ondrejov_period}, and it has a diameter of $1.88 \pm 0.01$ km as determined by NEOWISE \citep{2003UC20_diameter}. It had a close approach to the Earth of 0.035\,AU on 04 November /2023 \citep{small_body_database_lookup}.

\begin{figure}[H]
\includegraphics[width=0.6\textwidth]{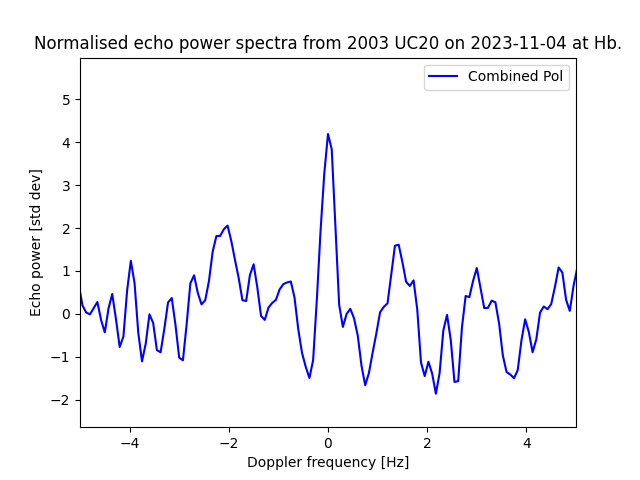}
\caption{Sum of spectra from the bistatic radar observation of 2003 UC20 from Hb with DSS-43 transmitting on 04 November 2023. Net length of scan is 201 minutes, starting at 15:47 UTC. The frequency received at the zero point is 7159.45 MHz, with a frequency resolution of 0.15 Hz. Radar echo detected with an SNR of 4-sigma when combining both linear polarisations.}\label{2003UC20_Hb}
\end{figure}

\textls[-20]{We detected the radar echo from an observation of 2003 UC20 on 04 November 2023 using the Hobart 12\,m antenna. This had the signal transmitted from DSS-43 to be centred at Hobart, so no Doppler compensation was required. The SNR was 4-sigma, as given in Figure \ref{2003UC20_Hb}, smaller than the 1994 PC1 detection. 2003 UC20 is both a smaller NEO and had a greater range than 1994 PC1 at the time of the observation, as given in Table \ref{Observations Table}, so this is not unexpected.}

Using Equation (\ref{eq:2}), the expected Doppler broadening was found to be 5.3 Hz. While the detected signal at Hobart was narrower, this could be explained by the rotation angle of the NEO relative to the antennas.

\subsection{2024 MK}\label{2024 MK}

\begin{figure}[H]
\includegraphics[width=0.6\textwidth]{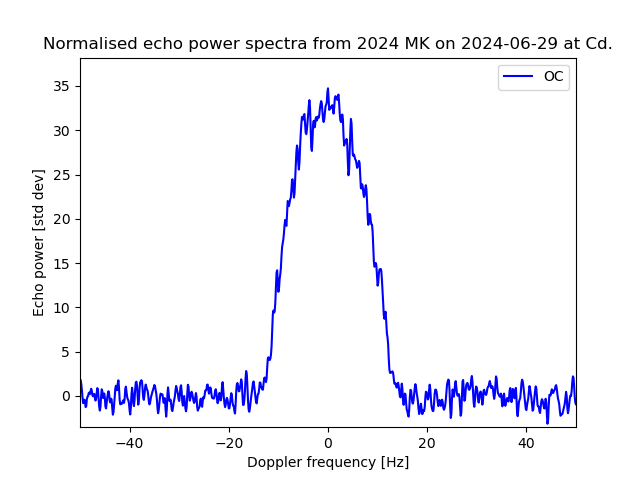}
\caption{Sum of spectra from the bistatic radar observation of 2024 MK from Ceduna with DSS-35 transmitting on 29 June 2024. The instantaneous radial velocity approximation was used for performing the Doppler compensation, and a frequency resolution of 0.25 Hz was used. A strong radar echo was detected with an SNR of 35-sigma in the opposite circular polarisation. Total scan length is 136 min.}\label{2024_MK_Cd}
\end{figure}

2024 MK was discovered by ATLAS-Sutherland (funded by NASA and located in South Africa) on 16 June 2024 and predicted to make a closest approach within 0.00197\,AU (0.77 lunar distances) on 29 June 2024. The object had an estimated diameter between 120 and 260 m. SHARP observations of 2024 MK were conducted during the DSS-43/ATCA time already scheduled for another approaching asteroid 2011 UL21 on 26 and 28 June and the Target of Opportunity DSS-35 time for 2024 MK on 29 June. The last experiment was conducted following the upgrades to Ceduna in May 2024 (see Section~\ref{Experiment setup}), so this station also participated. A strong radar echo was detected from Ceduna, Hobart, Katherine and Yarragadee, with an SNR of 35-sigma at Ceduna in the opposite circular polarisation, as given in Figure \ref{2024_MK_Cd}. 

\begin{figure}[H]
\includegraphics[width=0.6\textwidth]{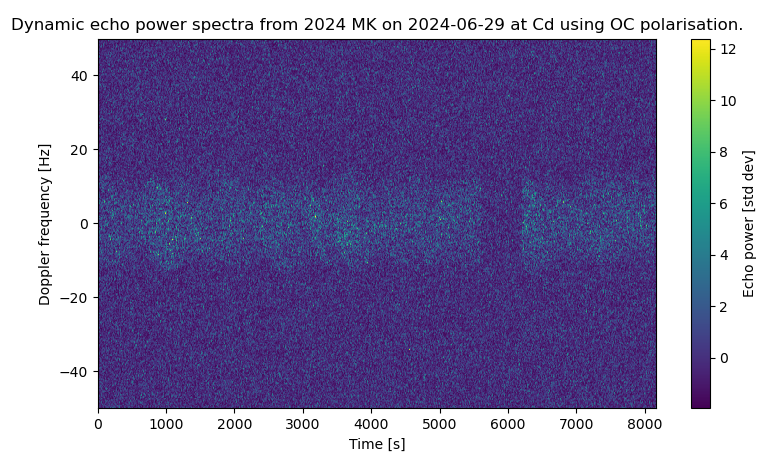}
\caption{Waterfall spectra plot from the bistatic radar observation of 2024 MK from Ceduna with DSS-35 transmitting on 29/06/2024, following compensation using the radial velocity approximation to Doppler.\label{2024_MK_Cd_Waterfall}}
\end{figure}

DSS-35 was transmitting, and the radial velocity approximation was used to perform the Doppler compensation, which can be seen to have negligible drift in Figure~\ref{2024_MK_Cd_Waterfall}. As there are no available values for its rotation period, its expected Doppler broadening cannot be estimated. However, we observe a data signature that resembles a rotational pattern with a period approximately between 1400\,s and 3100\,s using Equation (\ref{eq:2}).

\subsection{Optical Detection---2015 RN35}\label{Optical Detections}

\begin{figure}[H]
\includegraphics[width=0.55\textwidth]{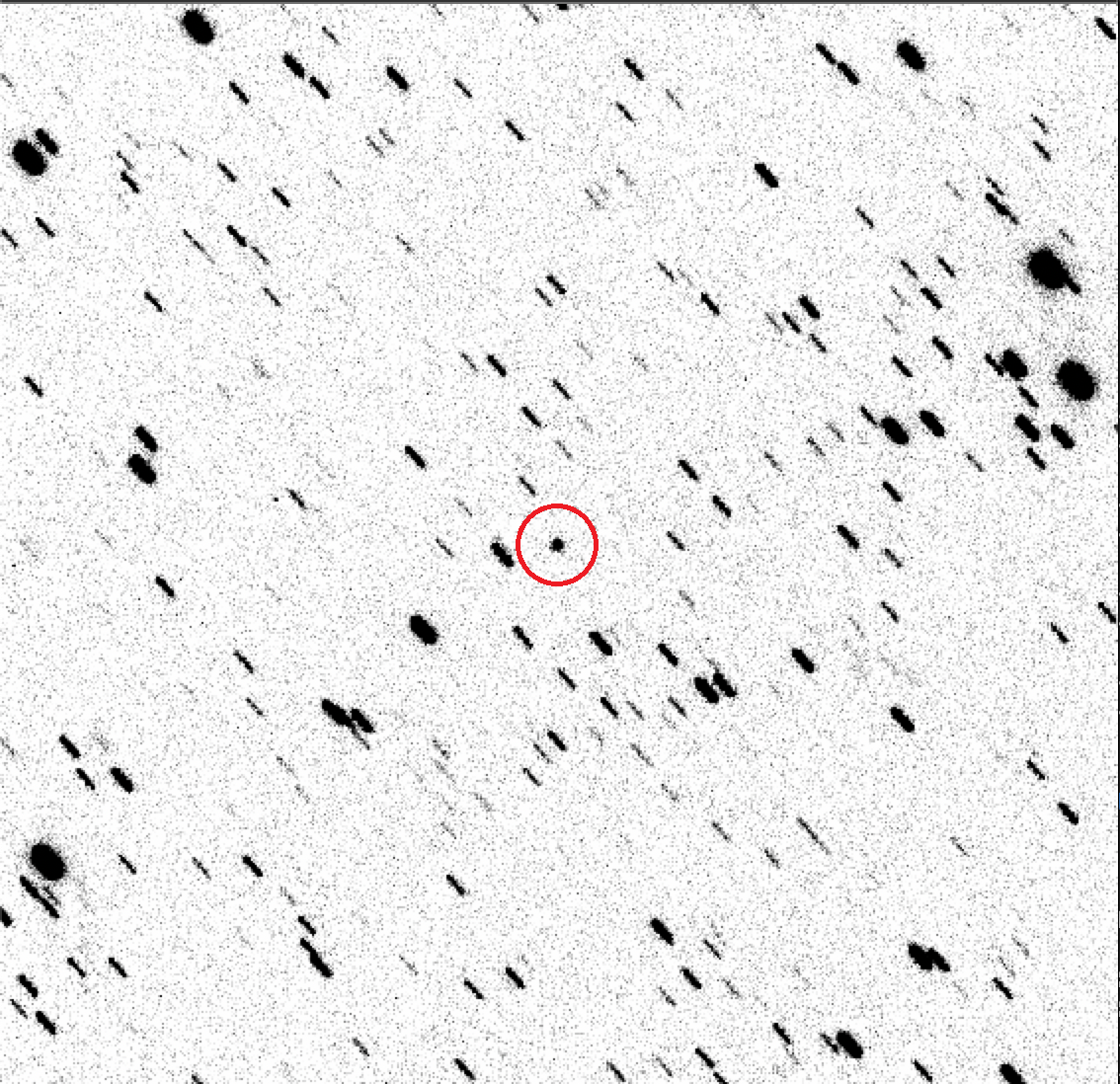}
\caption{60-s V band image of 2015 RN35 from the UTAS H50 telescope. The image is oriented such that north is up and east is to the left; the field of view is 10$^{\prime}$. 2015 RN35 is centred in the red circle.}\label{2015RN35_Optical}
\end{figure}

The University of Tasmania operates an 0.5 m Planewave CDK50 telescope (focal length 3454 mm) on an Alcor Direct Drive Nova120 equatorial mount with a maximum tracking rate of 20$^{\circ}$~s$^{-1}$. We are investigating the applications of simultaneous visible and radar tracking observations using this system, which is located 53~km north of the Hobart-26 site, at the UTAS Greenhill Observatory. On 18 December 2022 at UT 14:27, this telescope was used to observe 2015~RN35 as a capability demonstrator complementing the radar observations conducted on the previous day. Because accurate magnitude estimates were not available to the observer at the time, the observations were made at both the sidereal tracking rate and using the ephemeris tracking rate of the target in exposures of 60 s duration. A representative image is shown in Figure \ref{2015RN35_Optical}, in which the target is centred and the stars are trailed, confirming the expected motion relative to sidereal. The tracking rate was approximately 0.17 arcsec/sec in both RA and Declination across the observations. A total of 29 images were obtained in Bessell B and V filters, resulting in a broadband magnitude estimate of V$\approx$16.4. The results were used to validate expectations from our exposure time calculator, which will permit optimisation of exposure time for maximum resolution in future events.

\section{Discussion}\label{Discussion}

\subsection{Effectiveness of Data Processing Methods}\label{Effectiveness}

Both methods of using true Doppler from predict files and instantaneous radial velocity as an approximation have been shown to be useful for detection in compensating for the Doppler shift due to the relative motion of the stations and target. In particular, the method chosen had a negligible effect on the detection of 1994 PC1 as detailed in Section \ref{1994 PC1}. Furthermore, the strong detection of the Moon in Section \ref{Moon Detection} also reinforces the accuracy of the instantaneous radial velocity method. In instances where data integration over several hours is required to achieve a detection threshold exceeding 3 sigma, the utilisation of ``osod'' predictions becomes essential.

As illustrated in Equation~(\ref{eq:4}), the SNR of the received echo is a function of the gain on the receiver antenna and its system temperature. A direct comparison with the Goldstone and Arecibo observatories indicates that the 12\,m exhibits a performance deficit of 20.39 dB relative to Goldstone and 25.1 dB relative to Arecibo. We assume operations at 7.159 GHz and the same transmitted power. To compensate for the reduced sensitivity, a small network of antennas necessitates a longer integration time to integrate spectra. Specifically, to emulate the performance of a 70 m antenna such as DSS-43 or DSS-14, a small antenna would require at least 180 minutes of integrated spectra. When the SHARP bistatic configuration is considered in conjunction with Table 2 of~\citet{naidu_capabilities_2016}, i.e., DSS-43 transmitting at 20 kW and at 7.159 GHz, the relative sensitivity for UTAS small antennas is 0.003, while the Ceduna radio telescope exhibits a sensitivity of 0.008.

\subsection{UTAS Detections}\label{Detections}

As given in Table \ref{Observations Table}, there have been over 40 separate observation sessions but only three confirmed detections for the small antennas, 1994 PC1, 2003 UC20 and 2024 MK. The version of the radar range equation given in Equation (\ref{eq:4}) indicates that while many of these non-detections were expected, there were some that could have been. The strong detections of 2024 MK and the Moon indicate this is a signal strength problem rather than the Doppler compensation technique. Additionally, the current data processing pipeline implements a mixture of incoherent and coherent integration, with scans from 10 to 60 s coherently integrated and then each of these scans incoherently integrated to produce the final spectrum. Work is ongoing to optimise this balance to produce an optimal SNR.

Additionally, the bandwidth and frequency resolution for each observation could be optimised in the future based on the expected Doppler broadening. Particularly for weak signals, detectability could be improved through ensuring the entire radar echo is contained in a single frequency bin.

We are also working on data processing techniques to convert the linearly polarised signal of the broadband receivers to LCP and RCP in order to use the 12\,m antennas to analyse the SC and OC properties of the asteroid. To minimise potential sources of error in signal processing, a phase reference source can be utilised to calibrate the received power through radio interferometry. Data correlation provides estimates of delays caused by atmospheric conditions and facilitates antenna calibration for all the participant antennas in the observation. Interferometric calibration would be particularly beneficial for determining characteristics of NEO, such as spin-axis orientation, using VLBI techniques~\citep{Brozovic}.

\section{Conclusions}\label{Conclusion}

It has been demonstrated that small radio telescopes can be effectively used as receivers to partake in bistatic radar for strong targets, particularly in cases of large asteroids or those at very close proximity to the Earth. Successful observations of 1994 PC1, 2003 UC20, the Moon and 2024 MK were conducted using the Hobart and Katherine 12 m and the Ceduna 30\,m antennas.

The successful detections using small radio telescopes have prompted the upgrade of the larger antennas at the University of Tasmania. The Hobart-26 and Ceduna radio telescopes are now equipped with broader C-band receivers, operating at up to 7.3 GHz. This enhancement transforms them into valuable assets for future asteroid observations within the SHARP program. The capabilities of the Ceduna telescope have already been demonstrated with the 2024 MK experiment. Future observations will continue to test and calibrate both large antennas.

Future observations are planned to verify the performance of the Hobart-26 antenna and the possible use in regular campaigns. We adopted the open-source code developed specifically for support of planetary missions to near-Earth objects. The upgrades have eased the handling of different sorts of observations and scenarios. Furthermore, there exist multiple methods of applying Doppler corrections which effectively compensate for fixed transmission signals or those shifted to be centred at another station. {The methodology presented in this paper is applicable to any radio telescope, enhancing the sensor array for planetary radar. This improves detection and characterisation of current asteroids and increases our ability to assess and mitigate potential asteroid threats.}

New observations with the optical telescopes at the Greenhill Observatory (0.5\,m and 1.3\,m diameters) are also planned for late 2025 in conjunction with new SHARP radar observations. The aim is to improve the current data acquisition and signal processing pipeline and provide combined optical and radio NEO data.

\vspace{6pt} 

\authorcontributions{Conceptualisation, G.M.C. and S.H.; methodology, O.J.W. and G.M.C.; software, G.M.C.; resources, S.H., G.M.C., A.C., C.P. and J.S.; data curation, O.J.W., G.M.C. and N.S.; writing---original draft preparation, O.J.W.; writing---review and editing, O.J.W., G.M.C., S.H., P.E., E.K., J.G., N.S, L.B. and E.P. All authors have read and agreed to the published version of the manuscript.}

\funding{This research received no external funding.}

\dataavailability{The raw data supporting the conclusions of this article will be made available by the corresponding author on request.} 

\acknowledgments{We would like to acknowledge Bryn Emptage for conducting the optical observations at Bisdee Tier and subsequent data curation. Thank you to Patrick Yates-Jones for upgrading the tracking software for continuous tracking of asteroids. Thank you to the staff of the CDSCC for their assistance in conducting the SHARP observations. Thanks also to Matilda Downes Smolenski for assistance with producing Figure \ref{Bistatic_Radar_Diagram}. Part of this research was carried out at the Jet Propulsion Laboratory, California Institute of Technology under a contract with the National Aeronautics and Space Administration (80NM0018D0004).}

\conflictsofinterest{The authors declare no conflicts of interest.}


\abbreviations{Abbreviations}{
The following abbreviations are used in this manuscript:\\

\noindent 
\begin{tabular}{@{}ll}
NEA & near-Earth asteroid\\
NEO & near-Earth object\\
ATCA & Australia Telescope Compact Array\\
SNR & signal-to-noise ratio\\
SHARP & Southern Hemisphere Asteroid Research Program\\
CDSCC & Canberra Deep Space Communication Complex\\
DSS & Deep Space Station\\
Ti & Tidbinbilla 70\,m (DSS-43)\\
LD & Lunar Distances\\
Hb & Hobart-12 m\\
Ke & Katherine-12 m\\
Yg & Yarragadee-12 m\\
Ho & Hobart-26 m\\
Cd & Ceduna-30 m\\
LCP & left-circularly polarised\\
RCP & right-circularly polarised\\
GSSR & Goldstone Solar System Radar\\
VLBI & Very Long Baseline Interferometry\\
SDtracker & Spacecraft Doppler Tracker\\
SCtracker & Multi-tone spacecraft tracker\\
IF & intermediate frequency\\
AU & Astronomical Unit
\end{tabular}
}

\begin{adjustwidth}{-\extralength}{0cm}
\printendnotes[custom]

\reftitle{References} 


\PublishersNote{}
\end{adjustwidth}
\end{document}